\documentclass[11pt]{article}
\pdfoutput=1 % if your are submitting a pdflatex (i.e. if you have
\usepackage{graphicx}
\usepackage{calc}
\usepackage{rotating}
\usepackage[english]{babel}
\usepackage{graphicx}
\usepackage{subfig}
\usepackage{float}
\usepackage{amsmath}
\usepackage{amssymb}
\usepackage{amsthm}
\usepackage{latexsym}
\usepackage{dcolumn}
\usepackage{geometry}
\usepackage{soul}
\usepackage[table]{xcolor}
\usepackage{adjustbox} % Add this package
\usepackage{booktabs}
\usepackage{hyperref}
\usepackage{mciteplus}
\usepackage{slashed}
\usepackage{multirow}
\usepackage{hhline}
\usepackage{cite}
\def\gtwid{\mathrel{\raise.3ex\hbox{$>$\kern-.75em\lower1ex\hbox{$\sim
$}}}}
\def\vio{\mathrel{\hbox{$E$\kern-.60em\hbox{$/
$}}}}
\newcommand{\met}{\ensuremath{\slashed E_T}}

\newcommand{\ba}{\begin{eqnarray}}
\newcommand{\ea}{\end{eqnarray}}

\newcommand{\be}{\begin{equation}}
\newcommand{\ee}{\end{equation}}

\begin{document}

\begin{center}
{\Large \bf {Unveiling E$_6$SSM Scalar Diquarks at the HL-LHC}} \\%[0.45cm]
\vspace*{0.8cm}
{\large M.~Ali$^{a,b,c}$, S.~Khalil$^d$, S.~Moretti$^{e,f}$, S.~Munir$^{g,h}$, H.~Waltari$^e$} \\[0.25cm]
{\small \sl $^a$ East African Institute for Fundamental Research (ICTP-EAIFR), 
University of Rwanda, Kigali, Rwanda} \\[0.25cm]
{\small \sl $^b$ The Abdus Salam ICTP, Strada Costiera 11, 34135, Trieste, Italy} \\[0.25cm]
{\small \sl $^c$ SISSA International School for Advanced Studies, Via Bonomea 265, 34136, Trieste, Itlay} \\[0.25cm]
{\small \sl $^d$ Center for Fundamental Physics, Zewail City of Science and Technology, 6 October City, Giza, Egypt} \\ [0.25cm]
{\small \sl $^e$ Department of Physics and Astronomy,
Uppsala University, Box 516, SE-751 20 Uppsala, Sweden} \\ [0.25cm]
{\small \sl $^f$ School of Physics \& Astronomy,
University of Southampton, Southampton SO17 1BJ, UK} \\[0.25cm]
{\small \sl $^g$ Department of Physics, Faculty of Natural Sciences and Mathematics, St. Olaf College, Northfield, Minnesota 55057, USA} \\[0.25cm]
{\small \sl $^h$ Physics Department, Augustana University, Sioux Falls, South Dakota 57197, USA} \\[0.25cm]
{\small \url{mali@eaifr.org}, \url{skhalil@zewailcity.edu.eg}, \url{stefano.moretti@cern.ch}, \url{shoaib.munir@augie.edu}, \url{harri.waltari@physics.uu.se}}
\end{center}
\vspace*{0.4cm}

%%%%%%%%%%%%%%%%%%%%%%%%%%% Abstract %%%%%%%%%%%%%%%%%%%%%%%%%%%%%%

\begin{abstract}
\noindent
We investigate the phenomenology of scalar diquarks with sub-TeV masses within the framework of the $E_6$ Supersymmetric Standard Model (E$_6$SSM) at the Large Hadron Collider (LHC). The diquarks decay dominantly to third generation quarks and hence are mainly produced in pairs through QCD interactions. Focusing on the lightest of the six diquarks predicted by the model, we select some representative low masses for them in a parameter space region consistent with experimental constraints from direct searches for additional Higgs boson(s) and supersymmetry, as well as from flavor physics analyses. Using Monte Carlo simulations, we assess these benchmark points against the latest LHC results corresponding to an integrated luminosity of 140\,fb\(^{-1}\). We propose two analysis scenarios, one with a fully resolved final state and a second using boosted top jets, which is more suitable for the higher end of our mass range. We further evaluate the signal significance of the pair-production of these diquarks, when each of them decays into $tb$ pairs, at the $\sqrt{s}=13$\,TeV LHC Run 3 with design integrated luminosity of 300\,fb\(^{-1}\), and also at the 3000\,fb\(^{-1}\) High-Luminosity LHC (HL-LHC). Our analysis yields a statistical significance exceeding $3\sigma$ at the HL-LHC for diquark masses up to 1\,TeV, indicating promising prospects for their discovery. 

\end{abstract}

%%%%%%%%%%%%%%%%%%%%% Section 1 %%%%%%%%%%%%%%%%%%%%%%

\newpage
\section{Introduction}
\label{sec:intro}
The Standard Model (SM) has proven to be highly effective in elucidating most of the observations made by particle as well as astro-particle physics experiments thus far. Nevertheless, due to factors such as the hierarchy problem, the nature of the cold dark matter (CDM), CP violation (CPV), and the matter-antimatter asymmetry of the Universe, it is widely acknowledged that the SM represents only the low-energy limit of an extended theoretical framework. Since its launch, the LHC has been vastly anticipated to directly unveil new-physics (NP) particles. That has not occurred yet, possibly because the scale at which NP exists lies beyond the current reach of the LHC, and/or perhaps because NP is different from its most sought after types, that are predicted by some well-established and appealing frameworks. 

One potential way in which NP can manifest itself is a diquark -- a particle that only interacts with (and hence decays into) a pair of quarks. Such a particle appears in $E_{6}$ \cite{Hewett:1988xc} as well as Pati-Salam ($SU(4)_{C}\times SU(2)_{L} \times SU(2)_{R}$) \cite{Mohapatra:2007af} Grand Unified Theories (GUTs), and in $R$-parity violating SUSY models \cite{Barbier:2004ez}. It can contribute to $n-\bar{n}$ oscillations \cite{Baldes:2011mh, Mohapatra:1980qe, Babu:2008rq, Ajaib:2009fq, Gu:2011ff, Babu:2012vc}, and to various other similar processes \cite{Beaudry:2017gtw, Chen:2018stt, BhupalDev:2020zcy}. Diquarks have also been employed to address the strong CP problem, and they can potentially impact both indirect $(\epsilon_{K})$ and direct $({\epsilon}^{\prime} / \epsilon)$ CPV in kaons \cite{Barr:1986ky, Barr:1989fi}. Furthermore, diquark provided one of the plausible explanations of the forward-backward asymmetry observed in $t\bar{t}$ production at the Tevatron \cite{Shu:2009xf, Dorsner:2009mq,Dorsner:2010cu,Arhrib:2009hu,Ligeti:2011vt,Hagiwara:2012gy}.

Under $SU(3)_{C}$, quarks are $\bf 3$s, so that 
\be
\begin{aligned}
3 \otimes 3 &= 6 \oplus \bar{3},  \\
3 \otimes \bar{3} &= 8 \oplus 1,	
\end{aligned}
\ee
implying that the color scalar exotic states may transform as sextets, anti-triplets, octets or singlets.
%, octets, or singlets. 
The Lorentz-invariant interactions of these colored scalar exotic states with a Dirac spinor, $\psi = P_{L}\psi+P_{R}\psi$, with $P_{L,R}$ denoting the helicity projection operators, are given as
\be
\bar{\psi} \psi \phi\ + \bar{\psi^{c}}\psi  \Phi\,. 
\ee
The first term preserves $B+L$\footnote{Hereafter, $B(L)$ denotes the Baryon(Lepton) number, so that $B+L$ is effectively the global fermion number.} with $\phi$ being an octet or a singlet that only couples to spinors with different chiralities. The second term above clearly violates $B+L$ and $\Phi$, which transforms as a sextet or an anti-triplet, and only couples to two quarks with the same chirality, is identified as a diquark. Here $\Phi$ carries a net baryon number, $B=\pm \frac{2}{3}$, and it is a scalar if it transforms under $SU(2)_{L}$ either as a singlet or a triplet, but a vector if it transforms as a doublet. 

The Yukawa Lagrangian of a scalar $\Phi\equiv S$ is written as 
\be
{\cal L}_{S} = \lambda_{ij}^{(\alpha)} \bar{\psi}_{i}^{c} P_{L,R}\psi_{j} S_{\alpha} + h.c.\,,
\ee
while for the vector $\Phi\equiv V^\mu$ it is
\be
{\cal L}_{V} = \eta_{ij}^{(\beta)} \bar{\psi}_{i}^{c}\gamma_{\mu} P_{L,R}\psi_{j}V_{\beta}^{\mu} + h.c.
\ee
Here $\lambda$ and $\eta$ denote the couplings of $S$ and $A^\mu$, respectively, and the $i$ and $j$ indices reflect the quark flavors. In Tab. \ref{tb:DQT}, we show the classification of the ($\alpha=1-8$) scalar and ($\beta=1-4$) vector diquarks according to their charges under the SM gauge group. The sextets are symmetric in color, while the triplets are antisymmetric. The interchange of quarks plays a crucial role in determining the couplings of the diquark. If the diquark state is antisymmetric under the exchange of two quarks, it only couples to two quarks with different flavors, but not to same-flavor quark pairs. Thus, the antisymmetry of the diquark state under the SM gauge group implies an antisymmetry under flavor as well \cite{Giudice:2011ak}.

\begin{table}[tbp]
\centering
	\begin{tabular}{c|c|c}
		Diquark type & $SU(3)_{C}\times SU(2)_{L}\times U(1)_{Y}$ charges & Couplings    \\
		\hline
	$S_{1}$  &  (6, 3, +1/3)  &  $Q_{L}Q_{L}$   \\
	$S_{2}$  &  ($\bar{3}$, 3, +1/3)  &   $Q_{L}Q_{L}$   \\
	$S_{3}$  &  (6, 1, +1/3)  &   $Q_{L}Q_{L}, u_{R}u_{R}$  \\
	$S_{4}$  &  ($\bar{3}$, 1, +1/3)  &   $Q_{L}Q_{L},
	u_{R}u_{R}$   \\
	$S_{5}$  &  (6, 1, +4/3)  &  $u_{R}u_{R}$   \\
	$S_{6}$  &  ($\bar{3}$, 1, +4/3)  &  $u_{R}u_{R}$   \\
	$S_{7}$  &  (6, 1, \textendash2/3)  &  $d_{R}d_{R}$   \\
	$S_{8}$  &  ($\bar{3}$, 1, \textendash2/3)  &  $d_{R}d_{R}$  \\
	$V_{{1}}^{\mu}$   &   ($\bar{6}$, 2, +1/3)   &   $Q_{L}\gamma_{\mu}d_{R}$   \\
	$V_{{2}}^{\mu}$   &   (3, 2, +1/3) &    $Q_{L}\gamma_{\mu}d_{R}$  \\
	$V_{{3}}^{\mu}$   &   ($\bar{6}$, 2, \textendash 5/3) &   $Q_{L}\gamma_{\mu}u_{R}$   \\
	$V_{{4}}^{\mu}$   &  (3, 2, \textendash 5/3)   &   $Q_{L}\gamma_{\mu}u_{R}$   	
	\end{tabular}
	\caption{Charges of all possible scalar and vector diquarks under the SM gauge group, and the quark chiral pairs they couple to.  \label{tb:DQT}}
\end{table}
 
The prospects of the direct observation of a scalar diquark at the LHC have been previously investigated in \cite{Mohapatra:2007af, Tanaka:1991nr, Atag:1998xq, Cakir:2005iw, Chen:2008hh, Han:2009ya, Gogoladze:2010xd, Baldes:2011mh}, but such a particle  remains elusive to this day. These studies have, however, explored the production of a single diquark, while its pair-production remains largely unexplored. In this study we conduct a phenomenological analysis of the production of a pair of diquarks of the type $S_4$, naturally predicted by the E$_6$SSM \cite{King:2005jy}, which is one of the most well-founded $E_{6}$-inspired SUSY models. It offers a highly appealing theoretical resolution to the $\mu$ problem encountered in the Minimal supersymmetric Standard Model (MSSM), thanks to the $U(1)_{N}$ gauge symmetry that prohibits the presence of bilinear terms for the Higgs supermultiplets in the superpotential. 

As we discuss below, in the E$_{6}$ model the diquark couplings to quarks break an approximate symmetry and hence are small. Furthermore, the couplings to the third quark generation are the largest and hence the production of a single diquark through quark-quark annihilation is suppressed. The dominant production mode is then pair production via standard QCD interactions, which also reduces the dependence on model-dependent couplings. We test the consistency of the lightest diquark, referred to as $D$ here, predicted by the E$_6$SSM with the exclusion bounds from the LHC on an $S_4$ produced singly in association with a heavy charged Higgs boson, and decaying into a $tb$ pair. We then reconstruct a $D\bar{D}$ pair by applying a set of well-defined selection cuts on the semi-leptonic decays of $t\bar{t}$, in an attempt to establish the discovery prospects of $D$ states at the LHC. Our signal-to-background analysis predicts a sizeable statistical significance of their signature in the $t\bar{t}b\bar{b}$ channel at the HL-LHC.
 
The paper is structured as follows. Section \ref{sec:model} describes the E$_6$SSM, detailing in particular the characteristics of the scalar diquarks in it. In Section \ref{sec:LHC} we discuss our Monte Carlo ($MC$) analysis, performed to first evaluate the constraints on the $D$ mass from the most recent LHC searches of the $S_4$. Section \ref{sec:Rec} then outlines our event selection and methodology for reconstructing the $D\overline{D}$ pair. Finally, in Section \ref{sec:concl} we present our conclusions.

%%%%%%%%%%%%%%%%%%%%%%%%%%% Section 2 %%%%%%%%%%%%%%%%%%%%%%%%%

\section{\label{sec:model} The E$_6$SSM }

The $E_6$ group can be broken down to the SM along with one extra surviving $U(1)^\prime$ symmetry as
 \begin{align}
 	\begin{split}
        E_6 & {\xrightarrow{M_{\rm GUT}}}\,SO(10)\times U(1)_\psi \\
        &{\xrightarrow{M_*}}\,SU(5)\times U(1)_\chi\times U(1)_\psi\\
        &{\xrightarrow{M_{**}}}\,SU(3)_C\times SU(2)_L \times U(1)_Y\times U(1)^{\prime} \,,
 		\label{eq:breaking}
 	\end{split}
 	\nonumber
 \end{align}
%        E_6&\rightarrow SO(10)\times U(1)_\psi\\
% 		&\rightarrow SU(5)\times U(1)_\chi\times U(1)_\psi\\
% 		&\rightarrow SU(3)_C\times SU(2)_L \times U(1)_Y\times U(1)_N\,,
where $U(1)^{\prime}=U(1)_{\chi} \cos \theta + U(1)_{\psi}\sin \theta$. The E$_6$SSM is obtained for $\tan \theta=\sqrt{15}$, and thus implies $U(1)'\equiv U(1)_N$, under which the right-handed neutrinos transform trivially. In this SUSY model, the cancellation of anomalies occurs automatically if the particle spectrum at low energy includes three complete 27-dimensional representations of $E_{6}$, which is decomposed under  $SU(5) \times U(1)_N$ as 
\ba
27_i \rightarrow \left(10, \frac{1}{\sqrt{40}}\right)_i + \left(\bar{5}, \frac{2}{\sqrt{40}}\right)_i + \left(\bar{5}, \frac{-3}{\sqrt{40}}\right)_i + \left(5, \frac{-2}{\sqrt{40}}\right)_i + \left(1, \frac{5}{\sqrt{40}}\right)_i + (1, 0)_{i}\,,
\ea
with $i=1,\,2,\,3$. Each $27$-plet (which is in fact a supermultiplet) consists of one generation of SM fermions (the first two terms on the right), up- and down-type Higgs doublets and color triplet scalars $D$ and $\overline{D}$ (terms 3 and 4), a SM-singlet field with a non-zero $U(1)_{N}$ charge (term 5), and a right-handed neutrino (term 6) -  along with their respective spin-1/2 superpartners. 

Excluding non-renormalizable interactions, the most general gauge invariant low energy superpotential of the E$_6$SSM can be formulated as
 \be
 W_{\rm{E}_6\rm{SSM}} = W_0 + W_1 + W_2\,,
 \ee    
 where
\ba 
\begin{array}{lll}
\label{eq:402}
W_0 = &  \lambda_i \Phi (H^d_i H^u_i) + \kappa_{i} \Phi_i (D_{i} \overline{D}_i) \\
   & +\, \dfrac{1}{2}M_{i}N_i^c N_i^c + h_{ij} N_i^c (H_u L_j) + W_{\rm MSSM}(\mu=0)\,,\\
W_{1}=& {g^{Q}_{ijk}D_{i}Q_{L_{j}}Q_{L_{k}}+g^{q}_{ijk}\overline{D}_{i}d^{c}_{R_{j}}u^{c}_{R_{k}}}\,,\\
W_{2}=& {g^{N}_{ijk}N^{c}_{i}D_{j}d^{c}_{R_{k}}+g^{E}_{ijk}e^{c}_{R_{i}}D_{j}u^{c}_{R_{k}}+g^{Q}_{ijk}Q_{L_{i}}L_{L_{j}}\overline{D}_{k}}\,,
\end{array}
\ea
with summation over repeated family indices ($i,\,j,\,k = 1,\,2,\,3$) implied. The supermultiplets in the above equation are the SM singlets $\Phi_i$,  Higgs doublets $H_{d_i}$ and $H_{u_i}$,  right-handed neutrinos $N_{i}^{c}$,  charged leptons $e_{i}^{c}$, up-type quarks $u_{i}^{c}$, and down-type quarks $d_{i}^{c}$, and the left-handed quark and charged-lepton doublets, $Q_{i}$ and $L_{i}$, respectively. 

The E$_6$SSM is constructed so as to adhere to certain discrete symmetries. The presence of $U(1)_{N}$ automatically results in the conservation of $Z_{2}^{M}=(-1)^{3(B-L)}$, also known as $R$-parity, which ensures the stability of the Lightest Supersymmetric Particle (LSP) as a CDM candidate. Furthermore, in order to differentiate between the active and inert generations of the SM-singlets and the Higgs doublets, while also minimizing non-diagonal flavor transitions due to the Higgs sector, the superpotential is assumed to have an $Z_{2}^{H}$ flavor symmetry \cite{King:2005my,King:2005jy}. Since it forbids operators that allow the lightest exotic (s)quark to decay, it can only be an approximate symmetry, and $Z_{2}^{H}$-violating couplings less than $10^{-4}$ yield sufficient suppression of flavor-changing processes. Under this symmetry all the matter supermultiplets except $\Phi\equiv \Phi_3$, $H_d\equiv H^d_{3}$, and $H_u\equiv H^u_{3}$ are odd~\cite{King:2005my,King:2005jy}. Thus, $H^d_{\alpha}$, $H^u_{\alpha}$, and $\Phi_{\alpha}$, with $\alpha=1,\,2$, do not acquire vacuum expectation values (VEVs), and only $H_u$, $H_d$, and $\Phi$ constitute the Higgs sector. The VEV of $\Phi$ breaks the $U(1)_{N}$ symmetry, and generates the mass of the corresponding $Z^\prime$ boson as well as the effective $\mu$--term. 

Finally, $W_1$ and $W_2$ in Eq. (\ref{eq:402}) contain terms involving colored exotic states, which violate $B$ and $L$ in addition to $Z_{2}^{H}$, and can lead to rapid proton decay. One cannot define $B$ and $L$ of $D_i$ and $\overline{D}_i$ so that the complete Lagrangian is invariant under both the $U(1)_B$ and $U(1)_L$ global symmetries. However, for consistency of the proton lifetime in the model with the current experimental limits, either a $Z_L^2$ or $Z^2_B$ discrete symmetry can be imposed. Invariance under an exact $Z^2_L$ symmetry, under which all the superfields except the lepton ones are even, forbids the Yukawa interactions in $W_2$. This reduces the E$_6$SSM superpotential to $W_{0}+W_{1}$ only, and $B$-conservation implies that the $S_4$-type scalars $D_{i}$ and $\overline{D}_i$ are diquarks, with baryon number $-2/3$ and $+2/3$, respectively. This is the E$_6$SSM variant under consideration here.\footnote{For $B=\pm\frac{1}{3}$ and $L=\pm 1$, $W_1$ is forbidden instead, and the E$_6$SSM superpotential is reduced to $W_{0}+W_2$. In this variant of the model, the exotic scalars are $S_1$-type leptoquarks, and their LHC phenomenology was studied recently in \cite{Ali:2023kss}.} The behavior of the fields under the assumed symmetries is summarized in Tab. \ref{Des_Symm} for ease of understanding. 

\begin{table}[tbp]
	\begin{center}
		\begin{tabular}{|c|c|c|c|c|}
			\hline
			\multicolumn{1}{|c|}{Fields} & $Z^{M}_{2}$ & $Z^{H}_{2}$ &$Z^{B}_{2}$ & $Z^{L}_{2}$\\
			\hline
			$\Phi_{\alpha}$,\,$H_{d\alpha}$,\,$H_{u\alpha}$ & $+$ & $-$ & $+$ & $+$ \\
			$\Phi$,\,$H_d$,\,$H_u$                          & $+$ & $+$ & $+$ & $+$ \\
			$Q_{L_{i}}$,\,$d^{c}_{R_{i}}$,\,$u^{c}_{R_{i}}$ & $-$ & $-$ & $+$ & $+$ \\
			$L_{L_{i}}$,\,$e^{c}_{R_{i}}$,\,$N^{c}_{i}$     & $-$ & $-$ & $+$ & $-$ \\
			$\overline{D}_{i}$,\,$D_{i}$                    & $+$ & $-$ & $+$ & $+$ \\
			\hline
		\end{tabular}
	\end{center}
	\caption{\label{Des_Symm} Field charges under the discrete symmetries of the E$_6$SSM.}
\end{table} 

Choosing the field basis of the E$_6$SSM such that the Yukawa couplings of $\Phi$ to $D_i$ and $\overline{D}_i$ are flavor-diagonal results in mixing between their scalar components from the same family only. The calculation of the diquark masses therefore reduces to diagonalization of three $2\times 2$ matrices
\begin{equation}
\label{eq:mass}
\begin{array}{c}
M^2(i)=\left(
\begin{array}{cc}
M_{11}^2(i) & \mu_{D_i} X_{D_i} \\
\mu_{D_i} X_{D_i}  & M_{22}^2(i) 
\end{array}
\right)\,,\\
{\rm with}~M_{11}^2(i)=m^2_{D_i}+\mu_{D_i}^2+ \Delta_{D}\,,~M_{22}^2(i)=m^2_{\overline{D}_i}+\mu_{D_i}^2+\Delta_{\overline{D}}\,,~X_{D_i}=A_{\kappa_i}-\dfrac{\lambda_i}{\sqrt{2}\varphi}v_d v_u\,,
\end{array}
\end{equation}
and $i=1,\,2,\,3$. Here $\varphi$, $v_d$ and $v_u$ are the VEVs of $\Phi$, $H_d$ and $H_u$, respectively, $m^2_{D_i}$ are the soft scalar masses of $D_i$, and $m^2_{\overline{D}_i}$ those of $\overline{D}_i$. These mass parameters break global SUSY conditions, along with $A_{\kappa_i}$, the trilinear parameter associated with the coupling $\kappa_i$. The  $\mu_{D_i}$s  above are the masses of the exotic fermionic counterparts of $D_i$, given (in the leading approximation) by
\begin{equation}
\mu_{D_i}=\dfrac{\kappa_i}{\sqrt{2}}\,\varphi\,,
\label{404}
\end{equation}
while $\Delta_{D}$ and $\Delta_{\overline{D}}$ are the $U(1)_N$ $D$--term contributions to the diquark masses, set by $M_{Z^{\prime}}^2$ as
\begin{equation}
\Delta_{D}\approx -\dfrac{1}{5}M_{Z'}^2\,,\qquad
\Delta_{\overline{D}}\approx -\dfrac{3}{10}M_{Z'}^2\,.
\label{406}
\end{equation}
The D-terms of the MSSM sfermions arising from the U$(1)_{N}$ group are proportional to $M_{Z'}^{2}$ and have a positive sign. Hence we expect the MSSM sfermions to be somewhat heavy, beyond the exclusion limits from the LHC. This heaviness also suppresses the contribution of the MSSM states to low-energy observables. Therefore the exotic states can be among the lightest BSM states of the model.

If we assume one of the two linear superpositions of the scalar components of $D_3$ and $\overline{D}_3$  supermultiplets (in analogy with the SUSY sector, and for index-alignment with the only active generation of the Higgs sector) to be the lightest $S_4$--type diquark, which we denote by $D$ for simplification, its mass is given by
\begin{equation}
\begin{array}{rcl}
m^2_D&=&\dfrac{1}{2}\Biggl[M_{11}^2(3) + \Delta_{11}(3) +  M_{22}^2(3) + \Delta_{22}(3)\\[4mm]
&-&\sqrt{\Biggl(M_{11}^2(3) + \Delta_{11}(3) - M_{22}^2(3) - \Delta_{22}(3)\Biggr)^2 + 4 \Biggl(\mu_{D_3} X_{D_3} + \Delta_{12}(3)\Biggr)^2} \Biggr]\,,
\end{array}
\label{407}
\end{equation}
where $\Delta_{lm}$ ($l,\,m=1,\,2)$ are the loop corrections to the four terms in the (symmetric) diquark mass matrix in Eq. (\ref{eq:mass}) for the third generation. 

\begin{figure}[tbp]
	\centering
	\subfloat[]{{\includegraphics[scale=0.65]{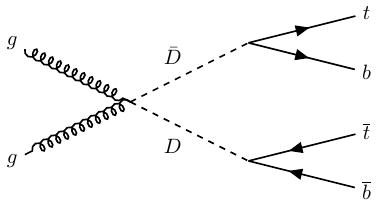} }}~~
	\qquad
	\subfloat[]{{\includegraphics[scale=0.65]{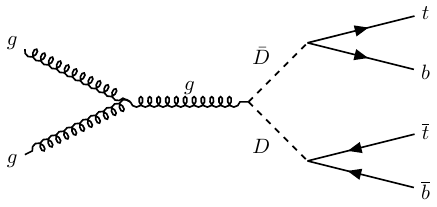} }}~~
	\qquad
	\subfloat[]{{\includegraphics[scale=0.65]{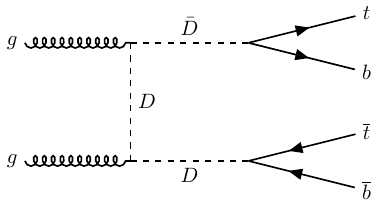} }} 
	\caption{LO pair-production of $S_4$-type diquarks ($D\bar{D}$) at the LHC.}
	\label{fig:FD} 
\end{figure}

 %%%%%%%%%%%%%%%%%%%%%%%%%%% Section 3 %%%%%%%%%%%%%%%%%%%%%%%%

\section{\label{sec:LHC} Constraints from the LHC}

Our analysis focuses on the pair production of a light ($\lesssim 1$~TeV) $D\bar{D}$ pair.\footnote{Recall that we have identified $D$ as the lightest of the six $S_4$-type diquark mass eigenstates predicted by the E$_6$SSM. Similarly, $\bar{D}$ simply implies the lighter scalar diquark originating from the 27-plet containing the third-generation antifermions. It is therefore a physical state with mass equal but electric charge exactly opposite to that of the $D$, and should not be confused for being the lightest of the $\overline{D}_i$ triplets appearing in the superpotential.} As noted in the previous section, the magnitude of mixing within each family of the diquark sector is governed by the parameters $X_{D_i}$ and $\mu_{D_i}$. While $\mu_{D_i}\propto \varphi$ is constrained by the collider limits on $M_{Z^\prime}\approx g_1^\prime \tilde{Q}_\Phi\varphi$, with $\tilde{Q}_\Phi$ being the $U(1)_N$ charge of $\Phi$, a large $X_{D_i}$ can still lead to mixing effects substantial enough that $D$ is one of the lightest SUSY particles, with a mass of the order of a few hundred GeV. Furthermore, if $D$ has $O(10^{-3})$ couplings only to the third-generation fermions (and vanishing ones to the others), it can have ${\rm BR}(D \to \bar{t} \bar{b}) = {\rm BR}(\bar{D} \to tb)\simeq 1$, where BR stands for Branching Ratio. 

The most dominant processes contributing to the production of $D\bar{D}$ pairs at the LHC are shown in Fig. \ref{fig:FD}. For our analysis, we computed the leading order (LO) (inclusive) cross section for $D\bar{D}$ production using {\tt MadGraph5\_aMC@NLO} \cite{Alwall:2014hca}, as 
 \ba
 gg \rightarrow D ~(\rightarrow \bar{t} \bar{b}) ~ \bar{D} ~(\rightarrow t b)\,,
 \label{DDbar}
 \ea
employing the {\tt NNPDF31\_lo\_as\_0118} parton distribution functions \cite{NNPDF:2017mvq} for the gluons, followed by
\ba
\bar{t} \rightarrow \ell^{-} \nu \bar{b}~~{\rm and}~~t \rightarrow j j b\, .
\label{FinalS}
\ea

Evidently, this final state can also originate from the production of a charged Higgs boson in association with $t\bar{b}$ pairs, and its total cross section is thus constrained by the corresponding searches in the ATLAS \cite{ATLAS:2021upq} and CMS \cite{CMS:2020imj} experiments. The ATLAS analysis pertains to a final state containing one charged lepton ($e$ or $\mu$) and jets, and the CMS analysis to an all-jet final state.

\begin{figure}[tbp]
	\centering
	\vspace{-1cm} 
	\includegraphics[scale=0.60]{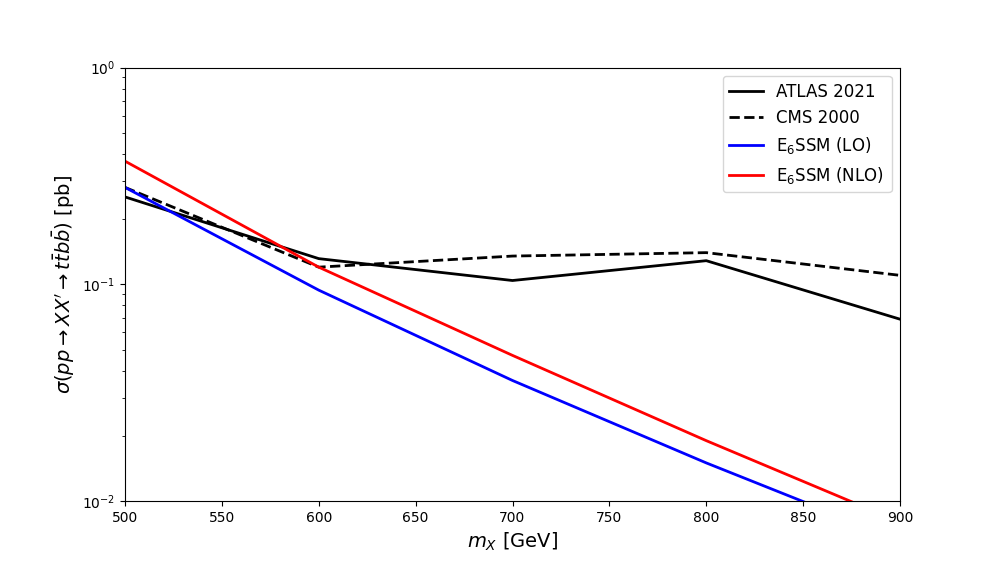} 
\caption{\label{fig:exc-limit} Exclusion contours in the $\{\sigma(t\bar{t}b\bar{b}),\,m_{H^\pm}\}$ plane from the ATLAS (solid black) and CMS (dashed black) data. The blue (red) line corresponds to the (N)LO cross section predicted for $t\bar{t}b\bar{b}$ production via $D\bar{D}$ in the E$_6$SSM. The intersection of this line with a given contour is interpreted here as the $m_D$ exclusion limit from the corresponding experiment.} 
\end{figure}

The Lagrangian of the phenomenological E$_6$SSM contains a number of additional parameters besides those of the Minimal Supersymmetric Standard Model (MSSM). To simplify our analysis, we set the values of the inert sector parameters such that it is decoupled from the low-energy model, which is effectively the MSSM augmented with the diquark sector. As mentioned earlier, the positive and large D-terms make the MSSM sfermions naturally heavy. Furthermore, the parameters related to the Higgs and neutralino sectors were fixed so as to yield a highly SM-like lightest Higgs boson, and a very Higgsino-like lightest neutralino with a sub-TeV mass, the consistency of which with the upper limit on the CDM relic abundance from Planck is therefore assured. The masses of the physical diquarks can be controlled, for a given $\mu_{D_i}$, by adjusting $m_{D_i}$ and/or $A_{\kappa_i}$, as noted earlier. For simplicity, we fixed $m_{D_i}^2= m_{\overline{D}_i}^2 = 2\times 10^6$\,GeV$^2$ and varied $A_{\kappa_i}$ slightly to obtain $m_D=$ 700\,GeV, 800\,GeV, and 900\,GeV. The particle spectrum for this parameter space point was obtained by incorporating the E$_6$MSSM into the public code \textsc{SPheno-v4.0.4} \cite{Porod:2003um,Porod:2011nf} using the Mathematica package \textsc{SARAH-v4.14.5} \cite{Staub:2008uz,Staub:2012pb,Staub:2013tta,Staub:2015kfa}. 

%\begin{gather}
%\tan\beta,\,g_N,\,\lambda,\,v_S,\,\lambda^T_3,\,M_i,\,M_{Q_i},\,M_{U_i},\,M^2_{D_i},%\,M^2_{\overline{D}_i},\,M_{L_i},\,M_{E_i},\,T_f(i,j)\,[\equiv Y_f(i,j) A_f(i,j)]%\nonumber\\
%M_{HuI}^2 (\alpha,\beta),\,M_{HdI}^2 (\alpha,\beta),\,E_{Mn}(i,j),\,E_\lambda(\alpha,\beta),\,E_\lambda^T(\alpha,\beta),\,g_{i,j,k}^Q,\,g_{i,j,k}^q,A_{\kappa_i},
%end{gather}
%where $i,\,j,\,k=1,\,2,\,3$, and $\alpha,\,\beta=1,\,2$, 
%for which $\{A_{\kappa_1},\,A_{\kappa_2},\,A_{\kappa_3}\}= \{0.276,\,0.32,\,0.35\}$\,GeV yielded $m_D=700$\,GeV. Increasing $A_{\kappa_1}$ to 0.284\,GeV and 0.291\,GeV, while setting $\{A_{\kappa_2},\,A_{\kappa_3}\} = \{0.3,\,0.33\}$\,GeV in either case,

In Fig. \ref{fig:exc-limit} we show the exclusion contours in the $\{\sigma(t\bar{t}b\bar{b}),\, m_{H^+}\}$ plane obtained from the $H^\pm b\bar{t}$ searches at both CMS and ATLAS. Of the two searches, the CMS one gives a stronger exclusion limit of $\sim 600$\,GeV on $m_D$ corresponding to the above-noted configuration of the E$_6$SSM parameters, for the $gg\to D\bar{D} \to t\bar{t}b\bar{b}$ cross section computed at the next-to-LO (NLO) at the $\sqrt{s}=13$\,TeV LHC Run III. For comparison, the LO cross section for the process is also shown in the figure. 

%%%%%%%%%%%%%%%%%%%%%%%%%%% Section 4 %%%%%%%%%%%%%%%%%%%%%%%%

\section{\label{sec:Rec} Event Analysis}

While limits from the LHC Run 2 analyses have been extracted using the full data sample (140 fb$^{-1}$), and the Run 3 analyses, which aim to exploit the full luminosity of 300 fb$^{-1}$, are still ongoing, there is a strong expectation that the integrated luminosity of 3000\,fb$^{-1}$ at the HL-LHC will greatly extend the scope of the searches for new physics phenomena. In order to estimate the prospects of the (HL-)LHC to observe $D\bar{D}$ pairs  in the $t\bar{t}b\bar{b}$ final state, we performed a comprehensive signal-to-background analysis. For maximal reconstruction efficiency of the signal events ($S$) in our analysis, we assume one of the two $W^\pm$ bosons (e.g., the one from $t\to Wb$) to decay hadronically, and the other one leptonically. Our final-state is thus comprised of four $b-$jets, two light jets, a single charged lepton, and missing transverse energy (\met) from the solitary neutrino. The background events ($B$) originate from 
\ba
p p \rightarrow t ~(\rightarrow j j b) ~\bar{t} ~(\rightarrow \ell^{-} \nu \bar{b})~ b \bar{b}~~{\rm and}~~
p p \rightarrow t~ (\rightarrow j j b)~ \bar{t} ~(\rightarrow \ell^{-} \nu \bar{b}) ~j j\,, 
\ea 
with the two light jets mis-tagged as $b-$jets in the latter.

To generate the signal and background events for our $MC$ analysis, we used (as mentioned) {\tt MadGraph5\_aMC@NLO} \cite{Alwall:2014hca} with {\tt NNPDF31\_lo\_as\_0118} parton distribution functions \cite{NNPDF:2017mvq}, in conjunction with {\tt Pythia-8.2} \cite{Sjostrand:2006za} for parton showering and hadronization, and with {\tt Delphes-3.4.2} \cite{deFavereau:2013fsa} for detector emulation. Jet reconstruction was performed using the anti-$k_t$ algorithm \cite{Cacciari:2008gp}, while {\tt MadAnalysis5} \cite{Conte:2012fm} was used for manipulating the MC data and plotting histograms. For effectively isolating the signal events from the background ones, we implemented the selection criteria detailed below.

%\subsection{\label{sec:ow}Low Mass Region}
\subsection{\label{sec:standard} Standard diquark reconstruction}

\begin{table}[tbp]
	\centering
	\begin{adjustbox}{max width=\textwidth}
		\begin{tabular}{l|c|c|c|c||c|c|c}
			\toprule
			& \multicolumn{4}{c||}{Signal $m_{D,\bar{D}}$} & \multicolumn{3}{c}{Background} \\
			 & 500\,GeV & 600\,GeV & 700\,GeV & 800\,GeV & $t\overline{t}b\overline{b}$ & $t\overline{t}c\overline{c}$ & $t\overline{t}jj$ \\
			\midrule
			$N_{MC}$  & 39200 & 12600 & 5040 & 2100 & 253400 & 214200 & 39480000 \\
			\midrule
			$N_{PS}$  & 929 & 324 & 134 & 56 & 2260 & 484 & 40199 \\
			$N_{M_{WH}}$  & 663 & 231 & 92 & 36 & 1635 & 345 & 28220 \\
			$N_{M_{tH}}$  & 384 & 126 & 47 & 17 & 1168 & 238 & 19135 \\
			$N_{M_{tL}}$  & 141 & 44 & 14 & 4 & 580 & 132 & 9239 \\
			$N_{p_T^b}$ & 21 & 11 & 5 & 2  & 10 & 1 & 154 \\
			\bottomrule
		\end{tabular}
	\end{adjustbox}
	\caption{Cut-flow for the four representative $m_D$ values, corresponding to our standard analysis, at the $\sqrt{s}=13$\,TeV LHC with an integrated luminosity of 140\, fb$^{-1}$.}
	\label{tb:cuts}
\end{table}

We used the `standard' cone size requirement, $R \leq 0.5$, for all the jets (including the $b-$jets) in the signal as well as the backgrounds. 
For the $b$-tagger we used a working point, where the $b-$tagging efficiency is $85\%$, the $c-$quark mistagging rate, $\epsilon_{c\rightarrow b}$, is 25\%, and the light--jet mistagging rate, $\epsilon_{u,\,d,\,s,\,g\rightarrow b}$, is 1\%. We then proceeded as follows. 
\begin{itemize}

\item $PS:$ Our set of preliminary selections included
\begin{center} 
$N(j)\geq 3,~ N(b)\geq 4,~N(\ell)=1,~p^{i}_{T} > 20\,\text{GeV}$, and $|\eta^{i}| < 2.5$. 
\end{center} 
Here $N(i)$ is the number of reconstructed objects of the type $i$, with $i=j,\,b,\,\ell$, while $p_T$ and $\eta$ are the transverse momentum and pseudorapidity, respectively, of a given object. 

\item $M_{WH}$: For the hadronic ($H$) $W^\pm$ decay, from all the possible pairings of the light jets, $j_{l}j_{m}$, with $l,m=1,\,2, \ldots N(j)$, with invariant masses lying within $20$\,GeV of the known $W$ mass, the one with $m_{j_{l}j_{m}}$ closest to $m_W$ was selected. 

\item $M_{tH}:$ Out of the four potential light--jet pair combinations with a $b-$jet, $jjb_{k}$ (for $k=1, ...4$), with invariant masses within $30$\,GeV of the known $t-$quark mass, the one with $m_{jjb}$ closest to $m_t$ was selected as the hadronic $t$ candidate. 

\item $M_{tL}$: For the leptonic ($L$) $W^\pm$, we combined $p_T^{\ell}$ with the \met, and require that $m_{\ell\nu}$ is exactly equal to $m_{W}$. This gives two solutions for $p^{\nu}_{z}$ \cite{Millar:2018dvh}, out of which we select that one the combination of which with $p_T^b$ gives the invariant mass closest to $m_t$, within a $30$\,GeV tolerance.

\item $p_T^b:$ Finally, we imposed $p_T^{b_{1,2}}>200$\,GeV on the leading and subleading ones among the remaining $b-$jets. We then reconstructed the diquark candidates by pairing the remaining two $b-$jets with the $t-$quark candidates, and choosing the pairing that gave the smallest $|m_{D,had}-m_{D,lep}|$, which is also required to be $<100$\,GeV, since the signal should correspond to a particular (though unknown) invariant mass.

\end{itemize}

\begin{figure*}[tbp]
	\centering
	\vspace{-1cm} 
	\includegraphics[scale=0.32]{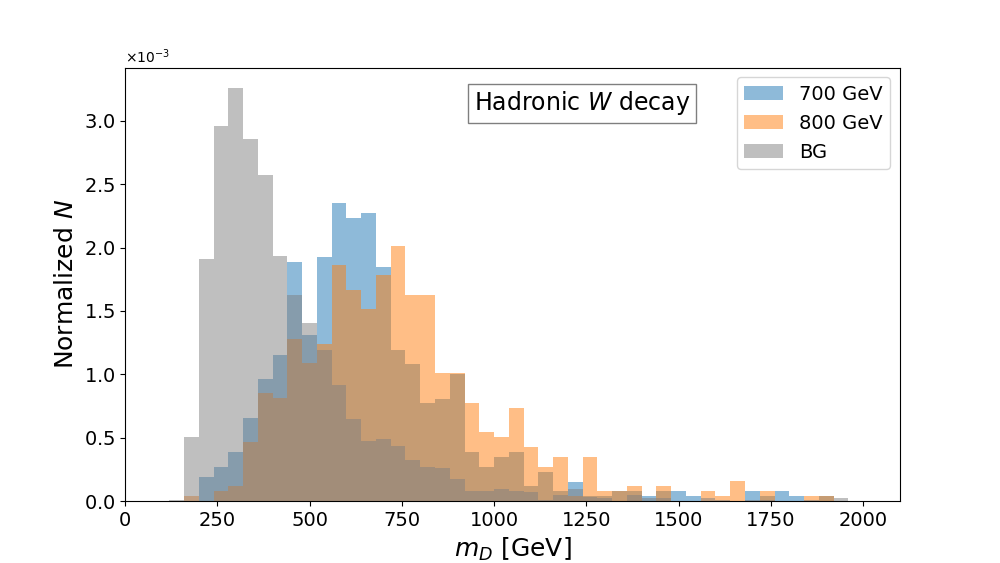}
	\includegraphics[scale=0.32]{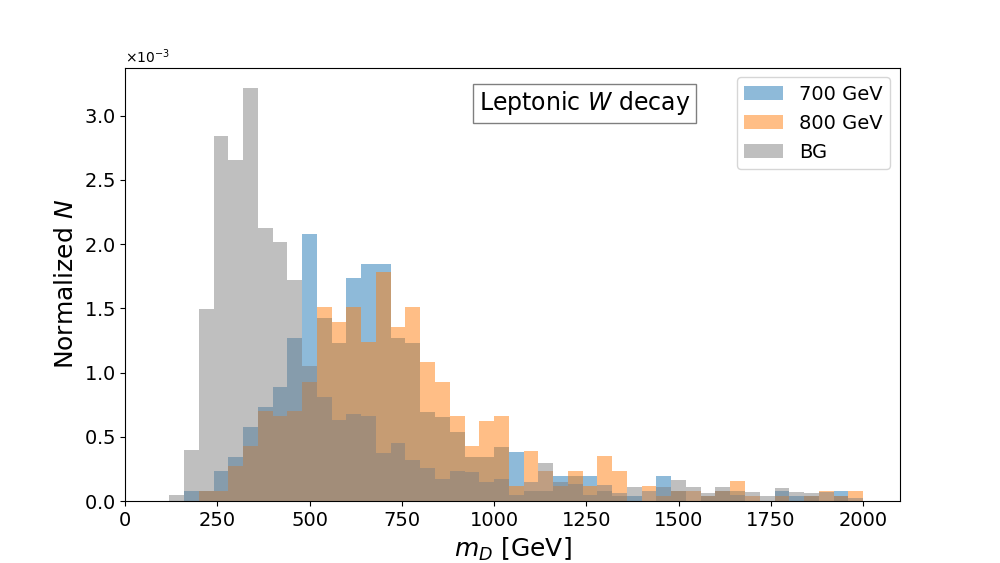}
	\includegraphics[scale=0.32]{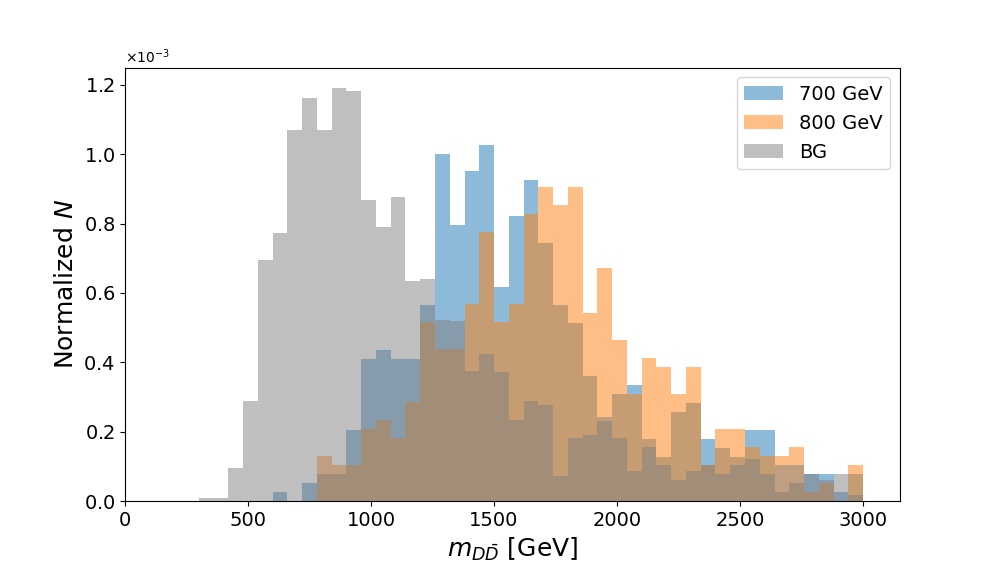}
	\caption{Signal and combined background event distributions (normalised to unity) for $m_{D}=700$ and 800\,GeV. Top row corresponds to the $D$ reconstructed from the hadronic (left) and leptonic (right) $W^\pm$, and bottom row to the $D\bar{D}$ pair.}
	\label{fig:stdrecon}
\end{figure*}

Tab. \ref{tb:cuts} displays the cut-flow for the signal corresponding to $m_D= \{500,\,600,\,700,\,800\}$\,GeV, and for the dominant backgrounds. The distributions of the $S$ and $B$ events for $m_D=700$ and 800\,GeV are shown in Fig. \ref{fig:stdrecon}, corresponding to the hadronic $W^\pm$ (top left panel) and the leptonic $W^\pm$ (top right panel). The bottom panel similarly illustrates the distributions for the invariant mass of the diquark pair, $m_{D\bar{D}}$. While the reconstructed masses are centered around the actual value, the distributions are quite widely spread, so a mass determination would be difficult.

We further calculated the signal significance, $N_S/\sqrt{N_{B}}$, where $N_S$ ($N_B$) is the number of events remaining for the signal ($t\overline{t}b\overline{b}$, $t\overline{t}c\overline{c}$, and $t\overline{t}jj$ backgrounds combined) after applying all the selection cuts noted above. These significances, computed using the $K$-factors for the signal as well as the backgrounds cross sections provided in Refs. \cite{Han:2009ya} and \cite{Choudhury:2024mox}, are shown in Tab. \ref{tb:Sig} assuming integrated luminosities of $140$, $300$, and $3000$ fb$^{-1}$ at the LHC, for $m_D=500,\,600,\,700$, and 800\,GeV. We notice a signal significance of $>5\sigma$ at the HL-LHC for a diquark with up to 600\,GeV mass. However, such low masses are already excluded by the ongoing LHC analyses, as noted in the previous section, while our standard event selection criteria fail to yield a statistically substantial isolation of the signal from the background for a higher $m_D$.

\begin{table}[tbp]
\centering
		\begin{tabular}{l|c|c|c|c}
			\toprule 
			 & $m_D = 500$\,GeV & $m_D = 600$\,GeV & $m_D = 700$\,GeV & $m_D = 800$\,GeV \\
			\midrule
LHC luminosity & \multicolumn{4}{|c}{$N_S/\sqrt{N_B}~(\sigma)$}  \\
			\midrule
			140/fb & 2.23 & 1.20 & 0.55 & 0.23 \\
			300/fb & 3.26 & 1.75 & 0.81 & 0.35 \\
			3000/fb & 10.33 & 5.54 & 2.58 & 1.11 \\
			\bottomrule
		\end{tabular}
\caption{\label{tb:Sig} Signal significances obtained with the standard event analysis for the current, design, and high luminosities at the LHC.}
\end{table}

\subsection{\label{sec:fatjet} Reconstruction of diquarks from fat jets}

The problem of the standard reconstruction method above is that the jets from a $D$ above a certain mass become highly collimated, and individual jets and leptons will can no longer be resolved as isolated objects in most cases. In such a scenario, the $t-$quarks can manifest as fat jets, implying that the hadrons (and possibly the leptons) are spread throughout a relatively wider cone, with no clear separation. Therefore, to improve the signal event selection for a relatively heavy $D$, we adopted an alternative strategy that incorporates boosted kinematics, wherein a $t-$quark is reconstructed as a single fat--jet, by setting $R\leq 0.8$ in the anti-$k_t$ algorithm. The rest of the analysis commenced as follows.

\begin{itemize}

\item $PS:$ The $b-$tagging and mistagging efficiencies, as well as the $p_T^i$ and $|\eta^i|$ requirements remained the same as in our standard analysis, but the other preliminary selections we imposed now read
\begin{center} 
	$N(j)\geq 2,~N(b)\geq 2,$ and $\sum H_{T}>1400$\,GeV,
\end{center} 
where $\sum H_{T}$ is the total hadronic transverse energy. Given that $m_{D}< 600$\,GeV is already excluded by existing LHC analyses, as discussed earlier, the signal events can be expected to have such a large hadronic activity.

\item $M_{tH}$: From these events we selected two fat--jets that had invariant masses lying in the $m_t \pm 30$\,GeV window as the two $t-$quark candidates. No $b$-tag requirement was imposed on the fat jets.

\item $p^b_T:$ These two fat--jets were then paired with the two leading $b-$jets, $b_1$ and $b_2$, each required to have $p_T > 150$\,GeV. The pairing that minimized $|m_{b_ifj_j} - m_{b_jfj_i}| < 150$\,GeV (where the subscript $fj$ implies a fat--jet, and $i,\,j=1,\,2$) was selected as the diquark pair candidate.

\end{itemize}

\begin{figure*}[tbp]
	\centering
	\includegraphics[scale=0.38]{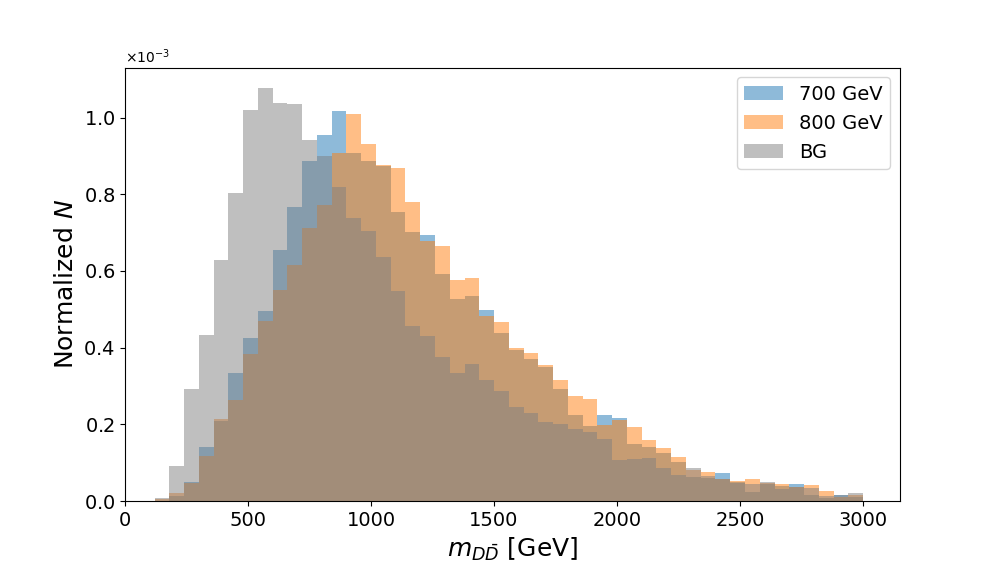}
	\caption{$D\bar{D}$ invariant mass distributions for the $m_{D}=700$ and 800\,GeV signals and the combined background, using the fat--jet selection criteria.}
	\label{fig:fjrecon}
\end{figure*}

The cut-flow for the events corresponding to $m_D= \{700,\,800,\,900,\,1000\}$\,GeV in the fat--jet analysis is given in Tab. \ref{tb:cuts1}, and the distributions of the $D\bar{D}$ invariant mass for the $m_D=700$ and 800\,GeV signals are shown in Fig. \ref{fig:fjrecon}.

\begin{table}[tbp]
	\centering
	\begin{adjustbox}{max width=\textwidth}
		\begin{tabular}{l|c|c|c|c||c|c}
			\toprule
			& \multicolumn{4}{c||}{Signal $m_{D,\bar{D}}$} & \multicolumn{2}{c}{Background} \\
			& 700\,GeV & 800\,GeV & 900\,GeV & 1000\,GeV & $t\overline{t}b\overline{b}$ & $t\overline{t}jj$ \\
			\midrule
			$N_{MC}$ & 5040 & 2100 & 840 & 420 & 253400 & 39480000 \\
			\midrule
			$N_{PS}$ & 3736 & 1518 & 585 & 282 & 172373  & 15462398 \\
			$N_{M_{tH}}$  & 117 & 59 & 27 & 16 & 340 & 18612 \\
			$N_{p^b_T}$ & 13 & 11 & 7 & 5 & 5  & 112 \\
			\bottomrule
		\end{tabular}
	\end{adjustbox}
	\caption{Cut-flow for the four representative $m_D$ values, corresponding to our fat--jet analysis, at the $\sqrt{s}=13$\,TeV LHC with an integrated luminosity of 140\, fb$^{-1}$.}
	\label{tb:cuts1}
\end{table}

\begin{figure*}[tbp]
	\centering
	\includegraphics[scale=0.45]{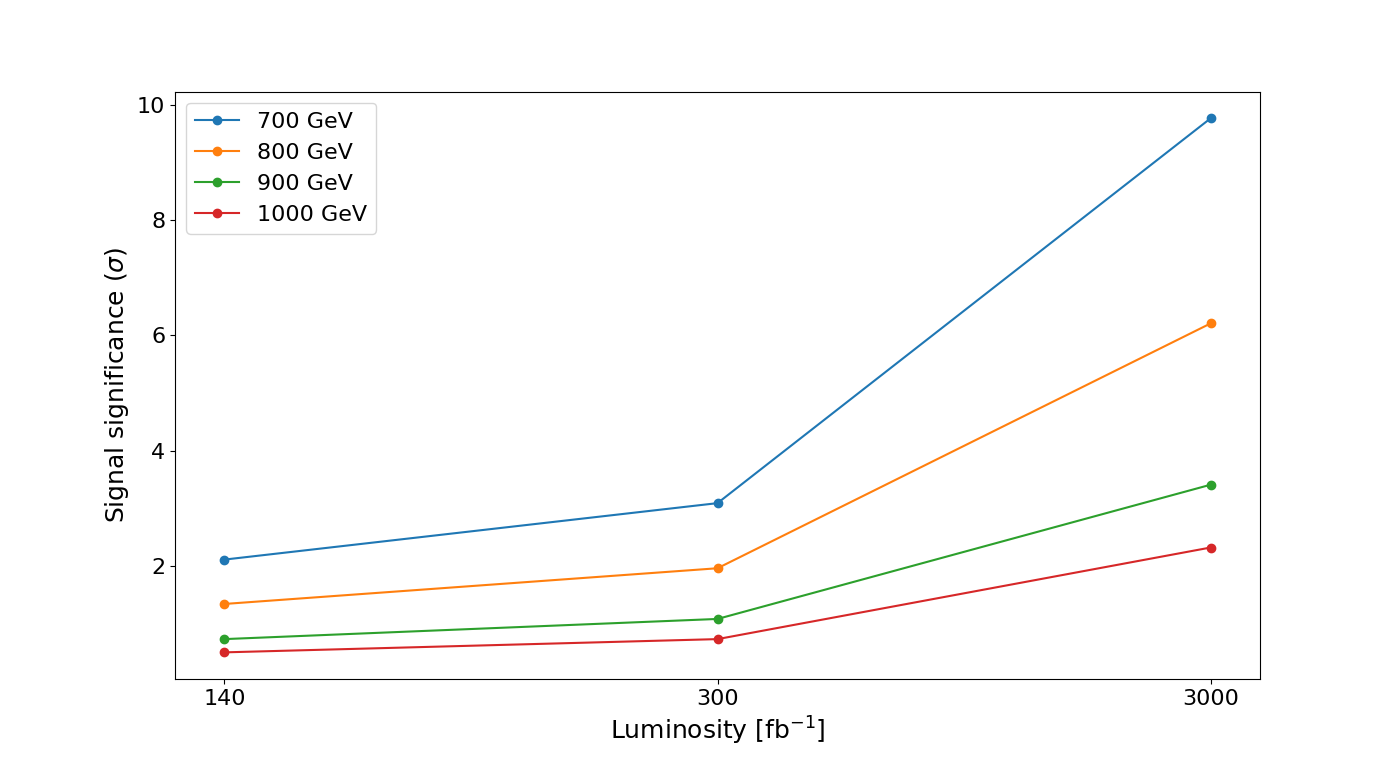}
	\caption{Estimated rise in the signal significances with our fat--jet analysis with increasing luminosity at the LHC, for pair-production of a $D$ with four representative mass values.}
	\label{fig:S_Pvalue}
\end{figure*}

According to Fig. \ref{fig:S_Pvalue}, the signal significances obtained using the fat--jet analysis are considerably enhanced compared to the ones obtained with the standard analysis, for $m_D=700$ and 800\,GeV. Even for $m_D=900$\,GeV this significance is above $3\sigma$, thus implying a fairly strong potential of the HL-LHC to discover a diquark with a sub-TeV mass in the final state considered here.

%%%%%%%%%%%%%%%%%%%%%%%%%%% Section 5 %%%%%%%%%%%%%%%%%%%%%%%%

\section{\label{sec:concl} Summary and Conclusions}

In this work, we have addressed a significant gap in the literature by performing a detailed phenomenological analysis of the lightest scalar diquark, denoted as \( D \), predicted by the E$_6$SSM. Focusing on the $\bar{t}\bar{b}$ ($tb$) decay channel of the $D$ ($\bar{D}$), we established the exclusion limits on its mass based on the most recent LHC data. We then determined the signal significance for the production of pairs of a fairly light (up to 1\,TeV) $D$ across various LHC scenarios, including its current and design luminosities, as well as the future HL-LHC with an integrated luminosity of 3000\,fb\(^{-1}\).  

Our analysis demonstrates that the HL-LHC could achieve a signal significance exceeding \( 3\sigma \) for \( D \) masses up to 1\,TeV, thus  indicating a strong potential for its discovery. Notably, our study goes beyond mere exclusion limits by presenting a robust detector-level analysis, underscoring the feasibility of detecting scalar diquarks of the E$_6$SSM framework in realistic experimental conditions.  

This study also highlights the importance of diquarks as indicators of beyond the SM physics, and establishes a basis for future research. Its extensions could investigate alternative decay channels, enhanced signal discrimination techniques, and explore wider parameter spaces, thereby advancing our understanding of exotic scalar states, and their relevance to high-energy physics.  

\section*{Acknowledgments}

The work of MA is supported by STEP and OEA from the Abdus Salam ICTP, Trieste, Italy. The work of SK is partially supported by the  Science, Technology \& Innovation Funding Authority (STDF) under grant number 48173. SM is supported in part through the NExT Institute and the STFC Consolidated Grant No. ST/L000296/1. HW is supported by Ruth and Nils-Erik Stenb\"ack's Foundation.

%\bibliographystyle{JHEP}
%\bibliography{E6SSMDQ}

\providecommand{\href}[2]{#2}\begingroup\raggedright\endgroup

\end{document}